\documentclass[a4paper,11pt]{article}
\newcommand{\be} {\begin{equation}}
\newcommand{\ee} {\end{equation}}
\newcommand{\ben} {\begin{equation*}}
\newcommand{\een} {\end{equation*}}
\newcommand{\bdm} {\begin{displaymath}}
\newcommand{\edm} {\end{displaymath}}
\newcommand{\bc} {\begin{center}}
\newcommand{\ec} {\end{center}}
\newcommand{\bea} {\begin{eqnarray}}
\newcommand{\eea} {\end{eqnarray}}
\newcommand{\bean}{\begin{eqnarray*}}
\newcommand{\eean}{\end{eqnarray*}}

\newcommand{\bfig} {\begin{figure}}
\newcommand{\efig} {\end{figure}}
\newcommand{\btab} {\begin{tabular}}
\newcommand{\etab} {\end{tabular}}

\usepackage{epsfig}
\usepackage{amssymb}
\usepackage{amsmath}

\def\NPB{{\em Nucl. Phys.} B}
\def\PLB{{\em Phys. Lett.}  B}

\def\PRD{{\em Phys. Rev.} D}

\begin{document}

\bc
{\bf \Large QUARK SPIN-FLIP IN POMERON EXCHANGE}
\footnote{Summary of talk given at the XIth International Conference on
Elastic and Diffractive Scattering, Blois 2005}\\
\vspace*{0.3in}
{\Large A. DONNACHIE}\\
\medskip
{\it \large School of Physics and Astronomy, University of Manchester,}\\
{\it \large Manchester M13 9PL, England}
\ec
\bigskip
\bc
{\bf Abstract}
\ec
\vspace*{-3mm}
{\small It has been shown \cite{AD05} that the energy-dependence of the 
reaction $\gamma p \to b_1(1235)$ requires a contribution from pomeron 
exchange. This necessitates spin-flip at the quark level as the transition is 
from a $^3S_1$ state to a $^1P_1$ state. The same mechanism occurs in the 
reaction $\pi p \to a_1(1260)p$, which is a $^1S_0$ to $^3P_1$ transition.}
\vspace{0.2in}

Preliminary H1 data \cite{H1} on $\gamma p \to (\omega\pi^0)X$ at $\langle W 
\rangle = 200$ GeV were provisionally interpreted as diffractive $b_1(1235)$ 
production:
\begin{equation}
\sigma(\gamma p \to b_1 X) = 790 \pm 200 \pm 200~~{\rm nb}.
\end{equation}
This interpretation seems unlikely as the transition $\gamma \to b_1(1235)$ 
does not satisfy the Gribov-Morrison rule \cite{Gri,Mor} $P_{\rm out} = 
(-1)^{\Delta J} P_{\rm in}$ and is from a $q\bar q$ spin-triplet state 
(photon) to a spin-singlet state $(b_1(1235))$, while it is 
well-established that helicity-flip amplitudes are small for pomeron exchange.

There is evidence from the Omega Photon Collaboration (CERN) \cite{CERN1} that 
the transition is not dominated by pomeron exchange as, for $20 \leq E_\gamma 
\leq 70$ GeV  ($\langle W \rangle = 8.6$ GeV) they find the energy dependence 
to be 
\begin{equation}
\sigma(E_\gamma) = \sigma(39)\Big(\frac{39}{E_\gamma}\Big)^\alpha,
\end{equation}
with 
\begin{equation}
\sigma(39) = 0.86 \pm 0.27 \mu{\rm b},~~~~\alpha = 0.6 \pm 0.2~.
\end{equation}
This implies a combination of Regge exchange $(\sim 1/E_\gamma)$ and pomeron 
exchange $(\sim E_\gamma^{2\epsilon},~~\epsilon \approx 0.08 - 0.1)$.

The CERN data are consistent with predominant $b_1(1235)$ production with 
$\sim 20\%$ $J^P = 1^-$ background, a result confirmed by SLAC \cite{SLAC} at 
$E_\gamma = 20$ GeV ($\langle W \rangle = 6.2$ GeV). If we assume 
non-interfering Regge exchange (responsible for producing the $b_1(1235)$) and 
pomeron exchange (responsiible for producing the $J^P = 1^-$
background) the cross section can be represented by
\begin{equation}
\sigma(s)=As^{2\epsilon}+Bs^{-2\eta}
\end{equation}
with $\epsilon=0.08,~\eta=0.4525,~A=0.107 \mu{\rm b},~B=29.15 \mu{\rm b}$. 
At $E_\gamma = 39$ GeV the pomeron-exchange contribution is $25\%$ of the 
total. Extrapolating to $W = 200$ GeV, this gives $584~{\rm nb}$ which becomes 
$\sim 730~{\rm nb}$ after including a factor for nucleon dissociation.

The $J^P = 1^-$ component can be estimated using simple VMD arguments:
\begin{equation}
\frac{d^2\sigma_{\gamma p \to V p}(s,m^2)}{dt~dm^2} =
\frac{\sigma_{e^+e^- \to V}(m^2)}{4\pi^2\alpha}
\frac{d\sigma_{Vp \to Vp}(s,m^2)}{dt}.
\end{equation}
Using the optical theorem to relate the amplitude at $t = 0$ to the total cross
section for $V p$ scattering and integrating over $t$ gives
\begin{equation}
\frac{d\sigma_{\gamma p \to V p}(s,m^2)}{dm} =
\frac{m\sigma_{e^+e^- \to V}(m^2)}{32\pi^3\alpha b}
(\sigma^{\rm Tot}_{V p \to V p}(s))^2,
\end{equation}
where $b \approx 5$ GeV$^{-2}$ is the slope of the near-forward differential 
cross section.

The Omega Photon Collaboration \cite{CERN2}compared $\gamma p \to (\pi^+\pi^-
\pi^+\pi^-)p$ with $e^+e^- \to \pi^+\pi^-\pi^+\pi^-$ over same energy and 
4-pion mass ranges as their $\omega\pi$ data, giving $\sigma^{\rm Tot}_{V p 
\to V p} = 16.7 \pm 3.4$ mb. It is now possible to predict $d\sigma/dm$ for 
the CERN and H1 data, given the data on $e^+e^- \to \omega\pi$. These data are 
shown in figure 1(a) and the prediction is compared with the CERN data in 
figure 1(b).

\begin{figure}[t]
\begin{center}
\begin{minipage}{65mm}
\epsfxsize65mm
\epsffile{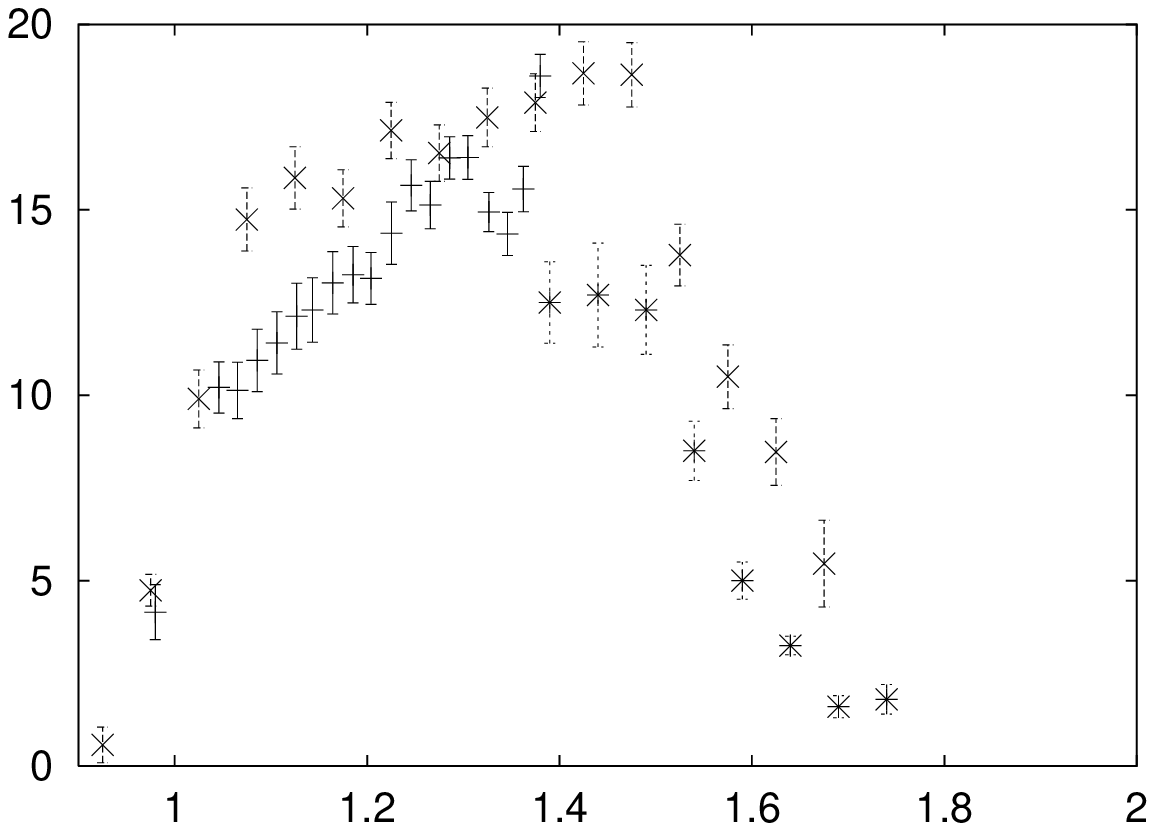}
\begin{picture}(0,0)
\setlength{\unitlength}{1mm}
\put(50,40){\small{(a)}}
\put(-5,40){\small{$\sigma$ (nb)}}
\put(30,-2){\small{$m$ (GeV)}}
\end{picture}
\end{minipage}
\hfill
\begin{minipage}{65mm}
\epsfxsize65mm
\epsffile{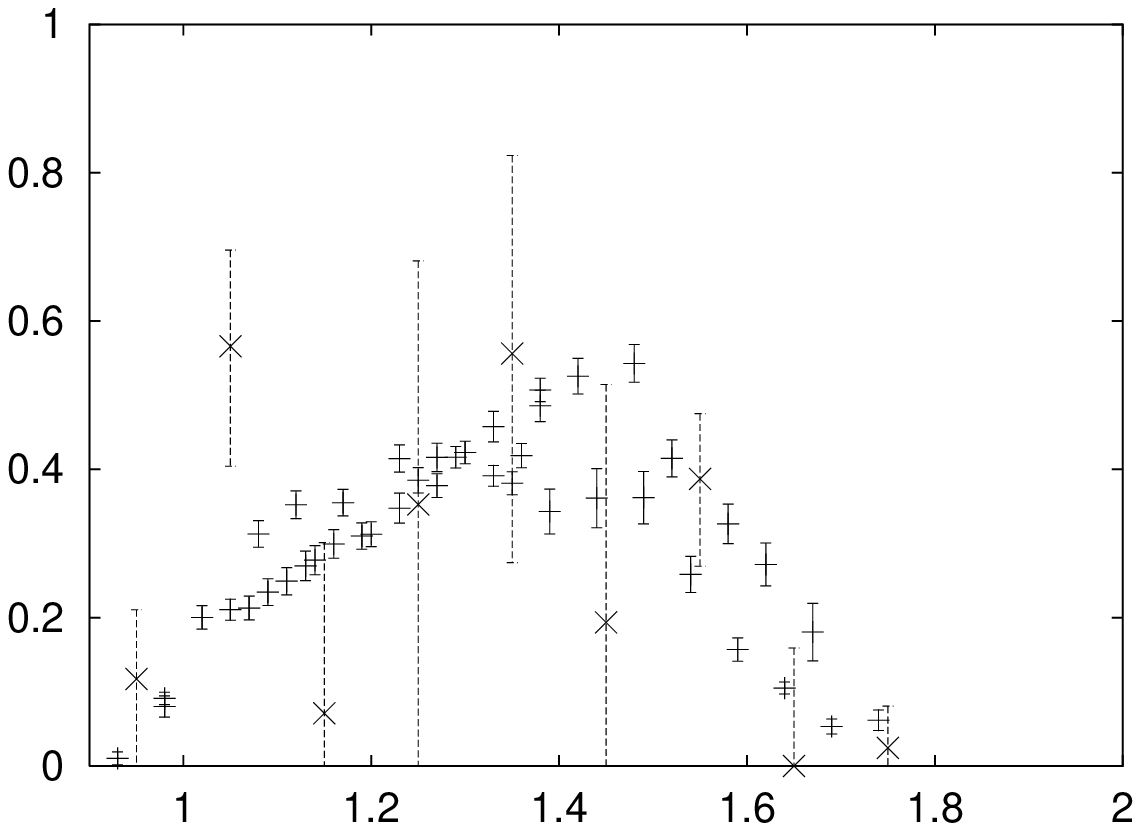}
\begin{picture}(0,0)
\setlength{\unitlength}{1mm}
\put(50,40){\small{(b)}}
\put(-5,40){\small{$\frac{d\sigma}{dm}$}}
\put(-8,35){\small{($\mu$b/GeV)}}
\put(30,-2){\small{$m$ (GeV)}}
\end{picture}
\end{minipage}
\end{center}
\caption{(a) The cross section for $e^+e^- \to \omega\pi$. The data are from 
Novosibirsk \cite{Novo} (horizontal bars), CLEO \cite{CLEO} (crosses) and
the DM2 Collaboration \cite{DM2} (stars)  (b) The $J^P = 1^-$ component of the
$\omega\pi$ mass distribution in the reaction $\gamma p \to \omega\pi p$ at 
$\sqrt{s}=8.5$ GeV. The data are from the Omega Photon Collaboration 
\cite{CERN1} (crosses) and from the application of vector meson dominance to 
the data in (a) (horizontal bars).}
\end{figure}

The predicted cross section is in good agreement with the H1 \cite{H1} at the 
upper end of the mass range, but there is an apparent excess of the data at 
the lower-mass end, as can be seen in figure 2. Is this indicative of some 
diffractive production of $b_1(1235)$ and, if so, is this reasonable?

\begin{figure}[t]
\begin{center}
\begin{minipage}{70mm}
\epsfxsize70mm
\epsffile{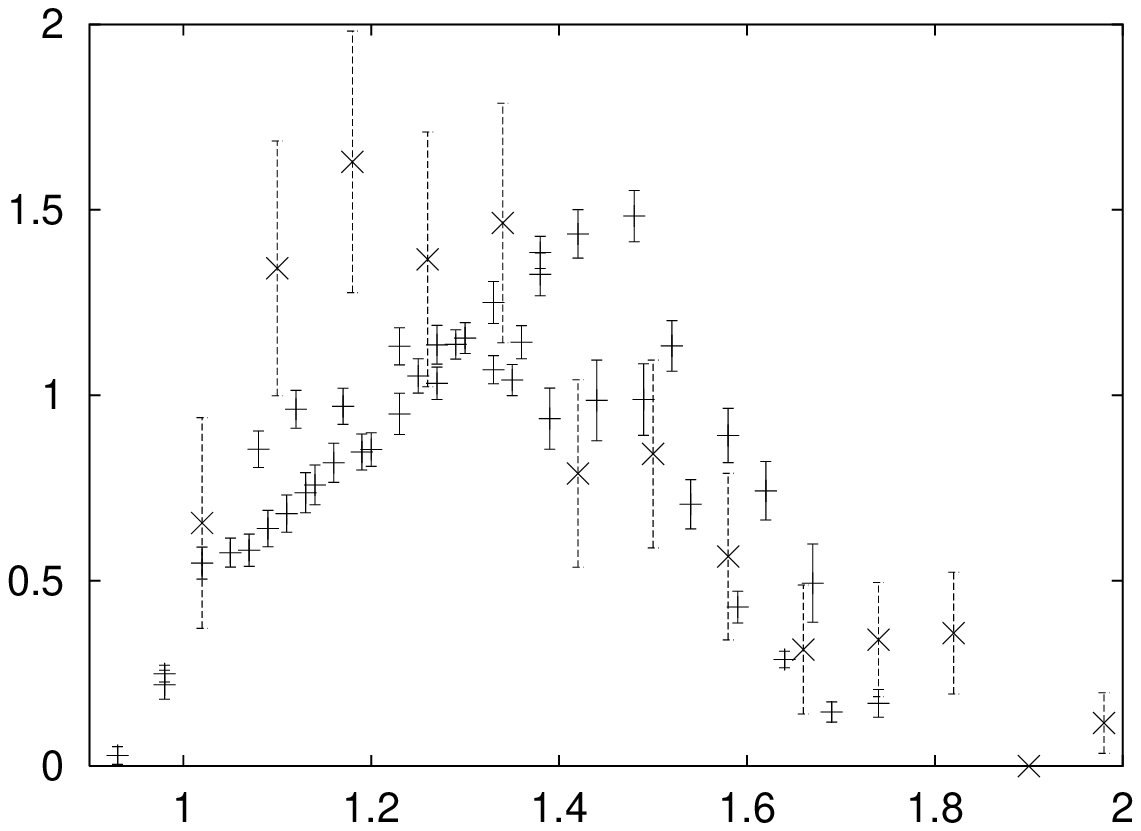}
\begin{picture}(0,0)
\setlength{\unitlength}{1mm}
\put(-5,40){\small{$\frac{d\sigma}{dm}$}}
\put(-9,35){\small{($\mu$b/GeV)}}
\put(30,-2){\small{$m$ (GeV)}}
\end{picture}
\end{minipage}
\caption{The $\omega\pi$ mass distribution in the reaction $\gamma p \to 
\omega\pi p$ at $\sqrt{s}=200$ GeV. The data (preliminary) are from the H1 
Collaboration \cite{H1} (crosses) and from the application of vector meson 
dominance to the data in figure 1(a) (horizontal bars).}
\end{center}
\end{figure}

An analogous reaction is $\pi^- p \to a_1(1260) p$. Although this does satisfy 
the Gribov-Morrison rule it requires spinflip at the quark level (singlet to 
triplet). Fitting the cross section data \cite{a1data} with a single effective 
power, $\sigma = As^\alpha$, gives $\alpha = -0.52$, close to the value found 
for $\gamma p \to b_1(1235)p$ implying the same interpretation of Regge plus 
pomeron exchange, but now we must allow for interference. A fit to the cross 
section data with
\begin{equation}
\sigma = As^{2\epsilon}+Bs^{\epsilon-\eta}+Cs^{-2\eta}
\end{equation}
gives $A = 7.87~~\mu{\rm b},~~B = 98.6~~\mu{\rm b},~~C = 1231~~\mu{\rm b}$.

Comparing the spin-flip pomeron-exchange contribution to the $\pi^- p \to 
a_1(1260) p$ cross section with the non-spin-flip pomeron-exchange contribution
to the $\pi p$ elastic scattering cross section shows the latter is a factor of
$\sim 130$ larger. A similar comparison of $\gamma p \to b_1(1235) p$ with 
$\gamma p \to \rho(770) p$ gives a ratio of the same order of magnitude. 

\begin{figure}[tbh]
\begin{center}
\begin{minipage}{70mm}
\epsfxsize70mm
\epsffile{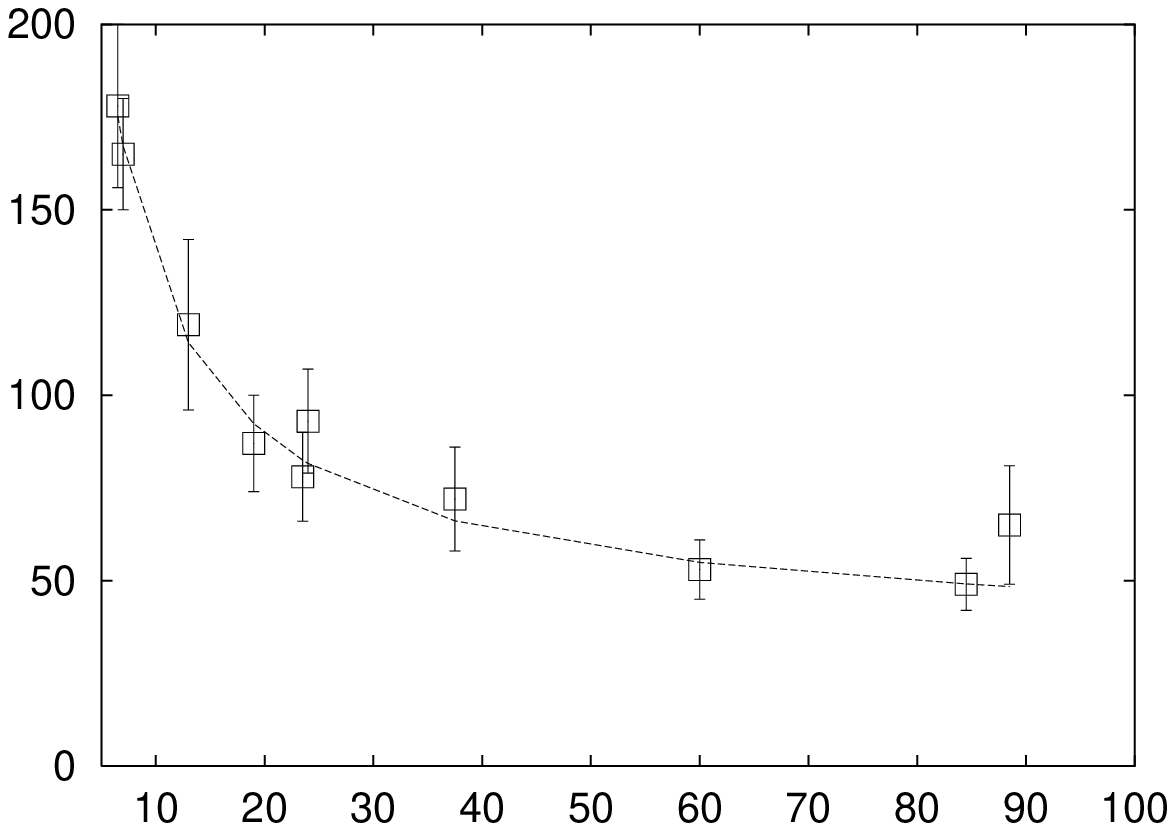}
\begin{picture}(0,0)
\setlength{\unitlength}{1mm}
\put(-5,42){\small{$\sigma$ ($\mu$b)}}
\put(30,-2){\small{$p_{\rm Lab}$ (GeV)}}
\end{picture}
\end{minipage}
\caption{The cross section for $\pi^- p \to a_1(1260) p$. The data are from the
ACCMOR Collaboration and the curve is the fit using the interfering Regge
plus pomeron parametrization}
\end{center}
\end{figure}

There are several reactions in which one would expect to see similar effects. 
Photoproduction of the isoscalar counterpart of the $b_(1235)$, namely the 
$h_1(1170)$, should occur at about $10\%$ of the photoproduction cross section
so would be of the order of 50 to 100 nb at HERA energies. The mechanism allows
diffractive photoproduction of the unconfirmed hidden-strangeness $h_1(1380)$,
which should occur at the level of $1\%$ of the $\phi$ photoproduction cross 
section, so we expect about 10nb at HERA energies. In the strange sector, the
$K_1(1270)$ and $K_1(1400)$ are nearly equal mixtures \cite{PDG} of the 
$K_{1A}$ ($1^3P_1$) and the $K_{1B}$ ($1^1P_1$) so we would expect the 
$1^3P_1$ component of the $K_1(1270)$ and $K_1(1400)$, as the analogue of the 
$a_1(1270)$, to be produced diffractively in high-energy $K p$ interactions.

Spin-flip coupling of the pomeron has been discussed extensively in the context
of proton-proton scattering at small $t$ \cite{MP02}, proton-proton scattering 
at large $t$ \cite{Gol02} and in vector-meson and $Q\bar Q$ production in deep 
inelastic scattering \cite{Gol03,GKP03}.

\end{document}